\documentclass[twoside,12pt]{article}
\usepackage{epsfig}
\usepackage{graphicx}

\def\lsim{\mathrel{\rlap{\lower4pt\hbox{\hskip1pt$\sim$}}
    \raise1pt\hbox{$<$}}}         
\def\gsim{\mathrel{\rlap{\lower4pt\hbox{\hskip1pt$\sim$}}
    \raise1pt\hbox{$>$}}}         

\def\Journal#1#2#3#4{{#1} {#2} (#4) #3 }

\def\NPB{{\em Nucl. Phys.} B}

\def\PLB{{\em Phys. Lett.} B}

\def\PRL{\em Phys. Rev. Lett.}

\def\PRD{{\em Phys. Rev.} D}
\def\PRC{{\em Phys. Rev.} C}

\newcommand{\be}{\begin{equation}}
\newcommand{\ee}{\end{equation}}
\newcommand{\bea}{\begin{eqnarray}}
\newcommand{\eea}{\end{eqnarray}}

\begin{document}

\title{ \vspace{1cm} Neutrino Oscillations}
\author{A.B.\ Balantekin,$^{1}$\\ 
W.C.\ Haxton$^2$\\ 
\\
$^1$Physics Department, University of Wisconsin, Madison WI 53706 USA\\
$^2$Department of Physics, University of California, and \\
Lawrence Berkeley National Laboratory,
Berkeley, CA 94720
}
\maketitle
\begin{abstract} This review summarizes recent experimental and theoretical progress in determining neutrino mixing angles and masses through neutrino oscillations. We describe 
the basic physics of oscillation phenomena in vacuum and matter, as well as
the status of solar, reactor, atmospheric, and accelerator neutrino experiments that probe these phenomena.
The results from current global analyses of neutrino parameters are given.  Future efforts that may improve the
precision with which these parameters are known or probe new aspects of the neutrino mixing matrix are described.
\end{abstract}
\section{Introduction}

The Standard Model (SM) of particle physics provides a renormalizable theory where charged fermions acquire 
Dirac masses through the Higgs mechanism. In
this model the neutrinos stand out, lacking charges or any other additively conserved quantum numbers to
distinguish particle and antiparticle, and consequently having the possibility of both Dirac and Majorana masses,
yet acquiring neither in the SM.
The structure of the SM -- the absence of right-handed fields necessary for Dirac masses and of dimensional
effective couplings necessary for Majorana masses -- eliminates neutrino mass terms.  
Consequently the discovery 15 years ago
of massive neutrinos supports the modern view of the SM as an effective theory, where for
example Majorana neutrino masses would arise from the SM's 
unique dimension-five operator, and thus can be considered the most natural correction to that model.
Hence neutrino masses represent
an important opportunity to look for extensions of the SM and to determine the scales of the new physics that
we now need to add to that model.  Experiment has confirmed that there is a hierarchy of neutrino masses,
just as there are differences between electron, muon, and tauon masses, though the overall scale of neutrino
masses is radically different from that of the charged fermions, below $\sim$ an eV rather than $\gsim$ MeV.  This has
implications for the scale of the new physics.  It has
also confirmed that there is flavor violation among the neutrinos -- the mass and weak interaction eigenstates
are different -- with the mixing angles being much larger than those found among the quarks. This has provided
us with a powerful tool for probing neutrino properties, oscillations that couple the flavors with significant amplitudes.

The discovery of neutrino masses and mixing, as well as the changes in the mixing that accompanies neutrino
propagation in matter, has stimulated a great deal of new experimental and theoretical activity.  Neutrino
oscillation experiments have
provided a reasonably clear picture of neutrino mixing angles and neutrino mass differences, though their are
important issues not yet unresolved.  Some of the remaining questions can be addressed with neutrino oscillations --
e.g., the mass hierarchy, the Dirac CP phase, and the possible existence of additional neutrino generations --
and some questions --  the absolute scale of neutrino mass, the nature of that mass
(its Dirac/Majorana character), and the Majorana CP phases -- must be addressed by other techniques.  This review summarizes the current status of
neutrino oscillation studies and prospects for further exploiting this probe to test neutrino properties.   

\section{Neutrino mixing and Oscillations}

The neutrino mixing matrix defines the relationship between the mass eigenstate and weak (flavor) eigenstate
bases,
\begin{equation}
\label{mix}
| \nu_{\rm flavor} \rangle = {\bf T} |\nu_{\rm mass} \rangle. 
\end{equation}
We adopt the following parametrization of the neutrino mixing matrix: 
\begin{equation}
\label{matrix}
 {\bf T} = {\bf T}^{(23)}{\bf T}^{(13)}{\bf T}^{(12)}  = 
\left(
\begin{array}{ccc}
 1 & 0  & 0  \\
  0 & C_{23}   & S_{23}  \\
 0 & -S_{23}  & C_{23}  
\end{array}
\right)
\left(
\begin{array}{ccc}
 C_{13} & 0  & S_{13} e^{-i\delta_{CP}}  \\
 0 & 1  & 0  \\
 - S_{13} e^{i \delta_{CP}} & 0  & C_{13}  
\end{array}
\right) 
\left(
\begin{array}{ccc}
 C_{12} & S_{12}  & 0  \\
 - S_{12} & C_{12}  & 0  \\
0  & 0  & 1  
\end{array}
\right)
\end{equation}
where $C_{ij} = \cos \theta_{ij}$, $S_{ij} = \sin \theta_{ij}$, and $\delta_{CP}$ is the CP-violating phase.  In writing Eq. (\ref{matrix}) we have omitted the two Majorana phases because these play no role in neutrino
oscillations. Note that the association of the CP-violating phase with $S_{13}$ is simply a parametrization choice. In fact the CP-violating difference of the neutrino survival probabilities 
\begin{equation}
\label{2}
P( \bar{\nu}_{\mu} \rightarrow \bar{\nu}_e) - P( \nu_{\mu} \rightarrow \nu_e), 
\end{equation}
is proportional to a product of the sines of all three mixing angles. 

Consider a particular flavor eigenstate, $| \nu_f \rangle = \sum_i T_{fi} |\nu_i \rangle$, created at some point in
time by weak interactions, where the indices $f$ and $i$ denote flavor and mass eigenstates respectively. 
The subsequent propagation of this state is controlled by the free Lagrangian and consequently by the neutrino
masses.  As the neutrino travels downstream from the source, the different mass eigenstates accumulate phases
that depend on their masses, with the differences in these phases governing the beating among the
mass eigenstates. If the initial state of definite flavor was created as a
 three-momentum eigenstate (a simplification, as one should employ wave packets), the amplitude for
 finding the state in its original flavor at time $t$ and distance $ct$ from the source is
\begin{equation}
\label{amp1}
A ( \nu_f \rightarrow \nu_f) = e^{i \vec{p} \cdot \vec{x}} \sum_i {\bf T}_{fi} e^{-iE_it} {\bf T}^{\dagger}_{if}.
\end{equation}
The time evolution of this state is governed by the $E_i = \sqrt{\vec{p}^2 + m_i^2}$, given that a momentum
eigenstate was assumed, which we expand for small neutrino masses as $p+ m_i^2/2p$.   The flavor of the state
thus evolves if some of the $m_i^2$ differ.  The electron neutrino survival probability downstream can be 
evaluated from Eq. (\ref{amp1}),
\begin{equation}
\label{ee}
P (\nu_e \rightarrow \nu_e) = 1 - \sin^2 2 \theta_{13} \left[ \cos^2 \theta_{12} \sin^2( \Delta_{31} L) + \sin^2 \theta_{12} \sin^2 (\Delta_{32} L) 
\right] - \cos^4 \theta_{13} \sin^2 2 \theta_{12} \sin^2 (\Delta_{21} L) 
\end{equation}
where $L$ is the distance traveled ($\hbar=c=1$) and
\begin{equation}
\label{9}
\Delta_{ij} \equiv \frac{\delta m_{ij}^2}{4E} = \frac{m_i^2-m_j^2}{4E}. 
\end{equation}
Similar probabilities can be evaluated for new flavors appearing downstream, with the probabilities summing to 1.
The mixing angles $\theta_{ij}$ and mass differences $\delta m_{ij}^2$ (include their signs) can thus be determined from the
variation of the neutrino oscillation pattern with distance and beam energy. If the case of oscillations among
three light neutrinos, only  two of the three $\delta m_{ij}$s are independent, as $\delta m_{12} +\delta m_{23}+ \delta m_{31} =0$. 

Just as photons traveling through matter propagate according to their index of refraction,
neutrinos traveling through matter acquire effective masses due to coherent forward scattering 
off surrounding particles \cite{Wolfenstein:1977ue}. 
The flavor dependence of this phenomenon -- ordinary matter contains electrons, but not muons or tauons --
can lead to striking effects, enhancing neutrino oscillation probabilities relative
to those seen in vacuum \cite{Mikheyev:1986gs}. The phenomenon is a familiar one, an adiabatic level
crossing, and is particularly dramatic for small mixing angles.  For example, an electron neutrino that might
be nearly coincident with a light neutrino mass eigenstate in vacuum ($\theta_{12}(\rho=0) \sim 0$)
would be coincident with the local heavy mass eigenstate at 
sufficiently high density ($\theta_{12}(\rho \rightarrow \infty) \rightarrow \pi/2$), as the effective mass contribution of
the matter overcomes the $\delta m_{21}^2$ difference that exists in vacuum.  For example, in the Sun the
electron neutrinos produced in the high-density solar core could be nearly coincident with the heavy-mass 
eigenstates.  If the neutrino then propagates adiabatically from high density to the vacuum that exists at the
solar surface -- this requires that the local oscillation length always be small compared to the scale height
of solar density changes -- the neutrino would then emerge into vacuum on the heavy-mass trajectory,
which would be nearly coincident with, say, the muon neutrino.
This phenomenon is known as the Mikheyev-Smirnov-Wolfenstein (MSW) effect.  
The oscillations that occur among three neutrino species in the presence of static matter 
is governed by the equation 
\begin{equation}
\label{msw}
i \frac{\partial}{\partial t} \left(\matrix{ \Psi_e\cr \Psi_{\mu} \cr
    \Psi_{\tau}} \right)  = \left[ {\bf T}^{(23)}{\bf T}^{(13)}{\bf T}^{(12)} \left(\matrix{
     E_1 & 0 & 0  \cr
     0 &  E_2  & 0 \cr
     0 & 0 &  E_3 }\right)
{\bf T}^{(12)\dagger}{\bf T}^{(13)\dagger} {\bf T}^{(23)\dagger} +
\left(\matrix{
     V_c+V_n & 0 & 0  \cr
     0 &  V_n  & 0 \cr
     0 & 0 &  V_n }\right)\right] \left(\matrix{ \Psi_e\cr \Psi_{\mu}
    \cr     \Psi_{\tau}} \right),
\end{equation}
where the Wolfenstein potentials in charge-neutral and unpolarized matter are
given by 
\begin{equation}
  \label{wolfen1}
V_c (\vec{x}) = \sqrt{2} G_F  N_e (\vec{x}) 
\end{equation}
for charged-current weak interactions and by
\begin{equation}
  \label{wolfen2}
V_n (\vec{x}) =  - \frac{1}{\sqrt{2}} G_F N_n(\vec{x}) 
\end{equation}
for neutral current interactions. In these equations $N_e$ and $N_n$ are the electron and neutron number densities of the medium, respectively. As an overall phase does not contribute to the interferences, we can take out a term proportional to the identity by defining 
$V_e = V_c +V_n$, $V_{\mu} = V_{\tau} = V_n$, and subtracting $V_{\mu} {\bf 1}$ from the second term in Eq. (\ref{msw}). 
Hence, the term $V_n$ drops out of the MSW equations for active neutrinos -- leaving $V_c$ to generate the
electron neutrino effective mass in the solar example given above. However, if sterile neutrinos are present, Eq. (\ref{msw}) must be expanded to include the additional neutrino species, and $V_n$ would then contribute to the
potential coupling active and sterile neutrinos. Also note that there is a 
SM loop correction \cite{Botella:1986wy}
\begin{equation}
\label{loop}
V_{\tau} - V_{\mu} = -  \frac{3 \sqrt{2} G_F\alpha}{\pi \sin^2 \theta_W}  \left( \frac{m_{\tau}}{m_W}\right)^2
\left\{ (N_p + N_n) \log \frac{m_{\tau}}{m_W} + \left( \frac{N_p}{2} + \frac{N_n}{3}  \right) \right\} , 
\end{equation}
where $N_p$ is the proton number density of the medium, $m_{\tau}$ is the tau lepton mass, $\theta_W$ is the Weinberg angle, and $m_W$ is the mass of the W-boson. 
This correction is negligible for the phenomena we consider here, but may be important in core-collapse supernovae. Similarly we will ignore interactions between neutrinos, which again play an important role in supernovae
due to the enormous neutrino densities produced by neutrino trapping \cite{bg}.

Often matter-enhanced neutrino oscillations, Eq. (\ref{msw}), are treated by retaining just two of the three  flavors. Indeed if the mixing angle $\theta_{13}$ were zero, Eq. (\ref{msw}) can be exactly reduced to a two-dimensional problem \cite{Balantekin:1999dx}. As two-flavor evolution is numerically simpler it is still widely employed. The contribution of the non-zero value of $\theta_{13}$ to the electron neutrino survival probability can be obtained using the formula 
\cite{Fogli:2001wi,Balantekin:2011ta}
\begin{equation}
\label{3x3}
P_{3\times3}( \nu_e \rightarrow  \nu_e) = \cos^4{\theta_{13}}  ( 1 - 4 \sin^2\theta_{13} \alpha) \> 
P_{2\times2}( \nu_e \rightarrow  \nu_e \>{\rm with}\> N_e
\cos^2{\theta_{13}})  + \sin^4{\theta_{13}}  (1+ 4 \cos^2 \theta_{13} \alpha)
\end{equation} 
where 
\begin{equation}
\label{alphadef}
\alpha = \frac{1}{\Delta_{32} + \Delta_{31}} \sqrt{2} G_F N_e (r=0) . 
\end{equation}
(We are envisioning an application to a case like the Sun, where spherical symmetry can be assumed,
and thus where densities are a function of the radial coordinate only.  Here $r=0$ is the solar center.)
This formula is obtained by expanding the full survival probability as a power series in $\sin \theta_{13}$. 
One can calculate the survival probability for the two-flavor case in Eq. (\ref{3x3}) by solving the equation 
\begin{equation}
\label{mswmod3}
i \frac{\partial}{\partial t} \left(\matrix{ \psi_e\cr
    \psi_x } \right) = 
\left(\matrix{
     \frac{1}{2} \tilde{V} - \Delta_{21} \cos 2 \theta_{12}&
    ~~~ \frac{1}{2} \Delta_{21} \sin 2 \theta_{12} \cr
   ~~~  \frac{1}{2} \Delta_{21} \sin 2 \theta_{12} & -
    \frac{1}{2}\tilde{V} + \Delta_{21} \cos 2 \theta_{12} }\right) 
\left(\matrix{ \psi_e \cr \psi_x }\right),
\end{equation}
where $\tilde{V} = \sqrt{2} G_F  N_e (r)  \cos^2\theta_{13}$ and $\psi_x = \cos{\theta_{23}} \psi_{\mu} -
\sin{\theta_{23}} \psi_{\tau}$. The MSW level crossing, where the adiabatic condition is most severe, is the point
where the diagonal terms are equal to zero. 

In some situations the matter basis simplifies the solutions;  by making the change of basis 
\begin{equation}
  \label{4}
  \left(\begin{array}{cc} \psi_1(r) \\ \\ \psi_2(r) \end{array}\right)
= \left(\begin{array}{cc} \cos{\theta_m(r)} & -\sin{\theta_m(r)} \\ \\
\sin{\theta_m(r)} & \cos{\theta_m(r)}
\end{array}\right)
\left(\begin{array}{cc} \psi_e(r) \\ \\ \psi_x(r)
\end{array}\right)\,,
\end{equation}
the flavor-basis Hamiltonian of Eq.~(\ref{mswmod3}) can be instantaneously
diagonalized. The cosine of this matter mixing angle is
\begin{equation}
\label{3}
\cos 2 \theta_m(r) = - \frac{(2 B E - \cos 2 \theta_{12})}{\sqrt{(2 B E - \cos 2 \theta_{12})^2 + \sin^2 2 \theta_{12}}}
\end{equation}
where 
\begin{equation}
\label{4a}
B = \sqrt{2} G_F N_e \cos^2 \theta_{13} / \delta m_{21}^2. 
\end{equation}
(This result corresponds to the more familiar two-flavor relationship between flavor
and local-mass eigenstates when $\theta_{13} \rightarrow 0$.)

Note that matter effects depend on the sign of $\delta m^2$, i.e., on the neutrino mass hierarchy. Furthermore, 
as the matter background is not CP-symmetric (no antiparticles), matter effects induce an effective CP violation. However, in the absence of sterile neutrino mixing, the CP-violating phase $\delta_{CP}$ does not impact neutrino oscillations even in the presence of a matter background 
\cite{Akhmedov2002,Balantekin:2007es}. 

\section{Measuring $\theta_{12}$: Solar and Reactor Neutrinos}

\begin{figure}
\begin{center}
\includegraphics[width=17cm]{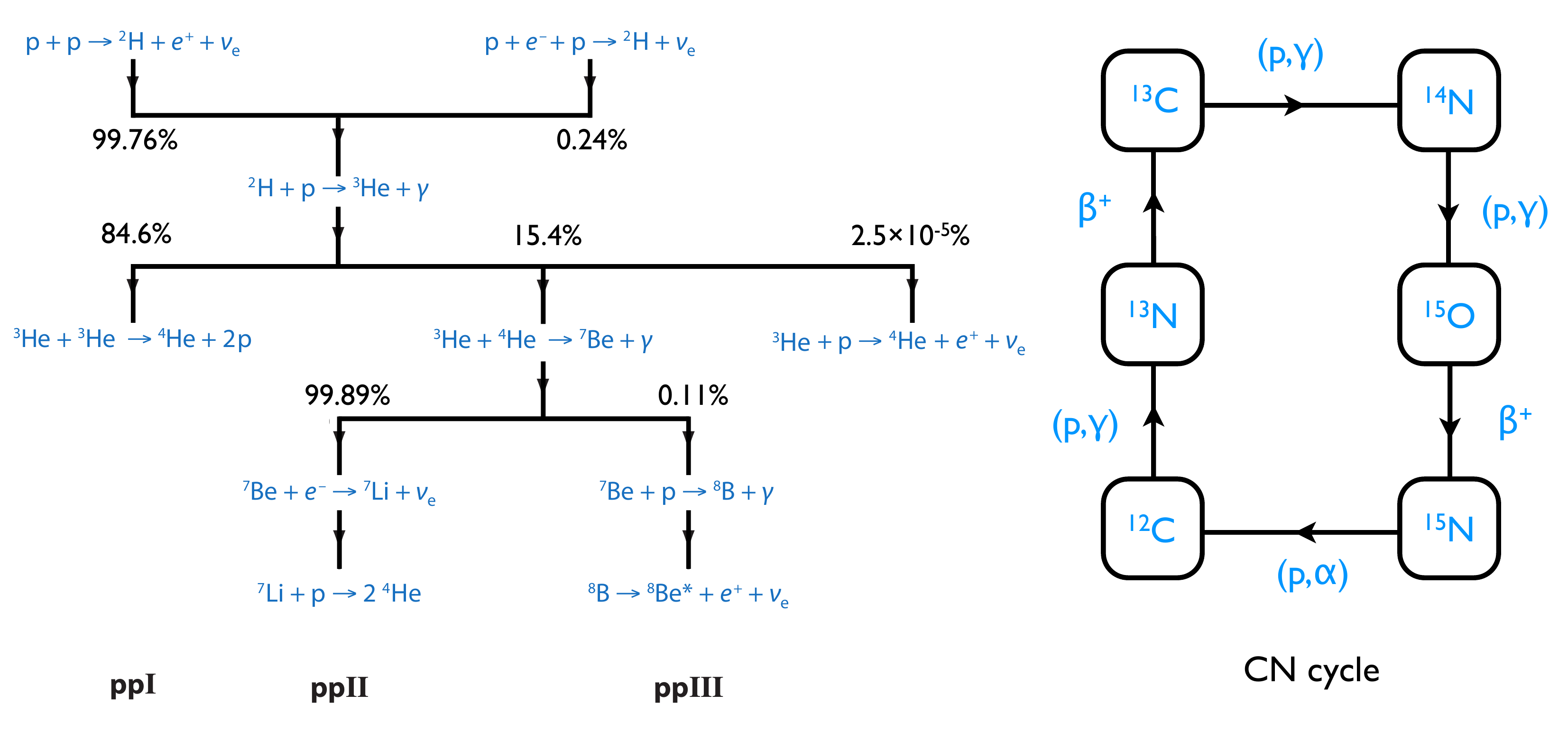}
\caption{The left frame shows the three principal cycles comprising the pp chain (ppI, ppII, and ppIII), the neutrinos 
sources associated with these cycles, and the theoretical
branching percentages defining the relative rates of competing reactions, taking
from the GS98-SFII SSM \cite{serenelli}.
Also shown is the minor branch ${}^3$He+p $\rightarrow$
${}^4$He+e$^+$+$\nu_e$, which generates the most
energetic solar neutrinos.  The right frame shows the CN I cycle, which produces
the $^{13}$N and $^{15}$O neutrinos. Figure from \cite{Robertson:2012ib}.}
\label{fig:cycles}
\end{center}
\end{figure}

The Sun and about 80\% of the visible stars produce their energy by the conversion of hydrogen to helium via
\begin{equation}
         2e^- + 4p \rightarrow \mathrm{^4He}  + 2 \nu_e + 26.73 \mathrm{~MeV}.
\end{equation}
Consequently stars, including our Sun, are prodigious sources of neutrinos. This conversion proceeds primarily through
the pp chain in lower mass, cooler stars like our Sun, and primarily through the CNO cycles in heavier mass, hotter
stars.   The pp chain, responsible for $\sim$ 99\% of solar energy production, is comprised of three
principal cycles, ppI, ppII, and ppIII, each of which is associated with a distinctive neutrino source.  The competition among these three cycles in very sensitive to temperature. The CN I cycle contribution to solar energy generation depends
on pre-existing C and N to catalyze reactions, and thus is sensitive to metallicity
as well as temperature.  Despite its minor role
in solar energy generation, the CN I  cycle produces neutrino fluxes from the $\beta$ decays of 
$^{13}$C and $^{15}$N that are potentially measurable.  The reaction chains are depicted in Fig. \ref{fig:cycles},
including pp-chain branching ratios from a recently updated Standard Solar Model (SSM), GS98-SFII \cite{serenelli}.
See Ref. \cite{Robertson:2012ib} for a recent comprehensive review of solar neutrino physics, to supplement the
summary provided here.

Pontecorvo first suggested that neutrinos might be detectable by radiochemical means in a large volume of
chlorine-bearing liquid (see the reproduction of his original paper in Ref. \cite{Pontecorvo:1946mv}).  This idea
was subsequently developed by L. Alvarez and later by R. Davis, Jr.  In 1964 Davis and his theoretical colleague
J. Bahcall argued in back-to-back Physical Review Letters \cite{Bahcall:1964gx,davis} that a
successful solar neutrino experiment could be done, given a sufficient detector volume and a deep underground location to
reduce cosmic-ray-associated backgrounds.  They gave as the scientific motivation for their proposal
``..to see into the interior of a star and thus verify directly the hypothesis of nuclear energy generation..."  First results
from the Cl detector were reported in 1968 \cite{Davis:1968cp}, and it soon became apparent that the measured solar neutrino
flux was about one third that predicted by the SSM of the time.  

The source of this
discrepancy was unclear for many years because of the large contribution of the high-energy
$^8$B neutrinos to the Cl capture rate: the flux of these neutrinos is
so sensitive to the solar core temperatures that a 5\% reduction in that temperature
would have brought theory into agreement with measurement.   The situation began to clarify 
around 1990, when three new experiments produced data.  The converted proton-decay
 water Cerenkov detector Kamiokande II reported  the results of 450 days of solar neutrino running in 1989,
 establishing that the $^8$B flux was $\sim$ 46\% the SSM model prediction \cite{Hirata:1989zj}, in approximate agreement with the Davis result.   The signal in this detector was the Cherenkov light produced by recoiling electrons
 after neutrino scattering.  This experiment was the first to detect neutrinos in real time and 
 the first to establish their solar origin by
 correlating the direction  of electron scattering events with the position of the Sun.  First results from SAGE and GALLEX,
 radiochemical detectors employing $^{71}$Ga as a target because of its low threshold and consequent
 sensitivity to pp solar neutrinos, were reported in 1991 and 1992, respectively.  The 1991 SAGE result,
 based on the first five extractions of the daughter nucleus $^{71}$Ge, was
 an upper bound on the counting rate of 79 SNU at 90\% c.l.\cite{Abazov1991}
 (SNU = solar neutrino unit, 10$^{-36}$ captures/target atom/sec).  The 1992 GALLEX result, based on an
 initial 14 extractions, was $83 \pm 19 ~\mathrm{(stat)} \pm 8~ \mathrm{(syst)}$ 
 SNU \cite{Anselmann1992}.  Both rates were uncomfortably
 close to the minimum astronomical value of 79 SNU \cite{Bahcall1989}, which assumes only a steady-state Sun
 and standard-model weak interaction physics.  
 GALLEX and its successor
 GNO operated until 2003 \cite{Hampel:1998xg,Altmann:2005ix}; SAGE remains an active experiment \cite{Abdurashitov2009}.  Because the chemistry of $^{71}$Ge 
 extraction from Ga is considerably more complicated than that of $^{37}$Ar extraction from Cl, efforts were made
 to calibrate the overall efficiency of both Ga experiments with
 intense $^{51}$Cr and $^{37}$Ar neutrino sources ($\sim$ 0.5 MCi) \cite{Abdurashitov:1998ne,Hampel:1997fc}.  Four such calibration experiments
 were performed, yielding a weighted average of measured to
 expected $^{71}$Ge of 0.87 $\pm$ 0.05 \cite{Abdurashitov2009}.   A counting rate of 66.1 $\pm$ 3.1 SNU \cite{Abdurashitov2009} was obtained by combining all results from
 SAGE, GALLEX, and GNO through 2009.  This rate can be compared to the
 GS98-SFII SSM prediction of 126.6 $\pm$ 4.2 SNU \cite{serenelli}.
 
 In the early 1990s, particularly with the uncertainties that accompanied the initial results from Kamioka II
 and the Ga experiments, no
 individual experiment 
 required a non-astrophysical solution of the solar neutrino problem.  But in aggregate, the Cl, Kamioka II, and 
 Ga experiments indicated a pattern of neutrino fluxes that was not compatible with any adjustment
 of the SSM.   This was pointed out in a series of papers by various authors as the data became
 more constraining, eventually establishing
with a high degree of confidence that the observed rates were
 inconsistent with the assumptions of a steady-state Sun producing undistorted fluxes of solar 
 neutrinos \cite{HL1994,P1995,WKG1993,HBL1994,HR1996}.



These early results combined with the spectacular success of the SSM in another context -- reproducing the
sound speed profile deduced from helioseismology to
sub-1\% accuracy throughout much of the solar interior \cite{Bahcall1989} -- 
made a particle-physics solution to the solar neutrino problem more likely.  This is turn helped 
motivate a new generation of spectacularly capable active detectors.  These experiments are
Super-Kamiokande, a 50-kiloton water Cherenkov detector from which the $^8$B flux was determined to better
than 3\%;  the Sudbury Neutrino Observatory (SNO), which employed a central 1-kiloton volume of heavy water
in order to measure three neutrino reaction channels with different sensitivities to flavor;  and Borexino,
which uses liquid scintillator, extending real-time, event-by-event counting to low energy fluxes
such as the $^7$Be and pep neutrinos.   Both Super-Kamiokande and Borexino continue to operate.

\begin{figure}
\begin{center}
\includegraphics[width=16cm]{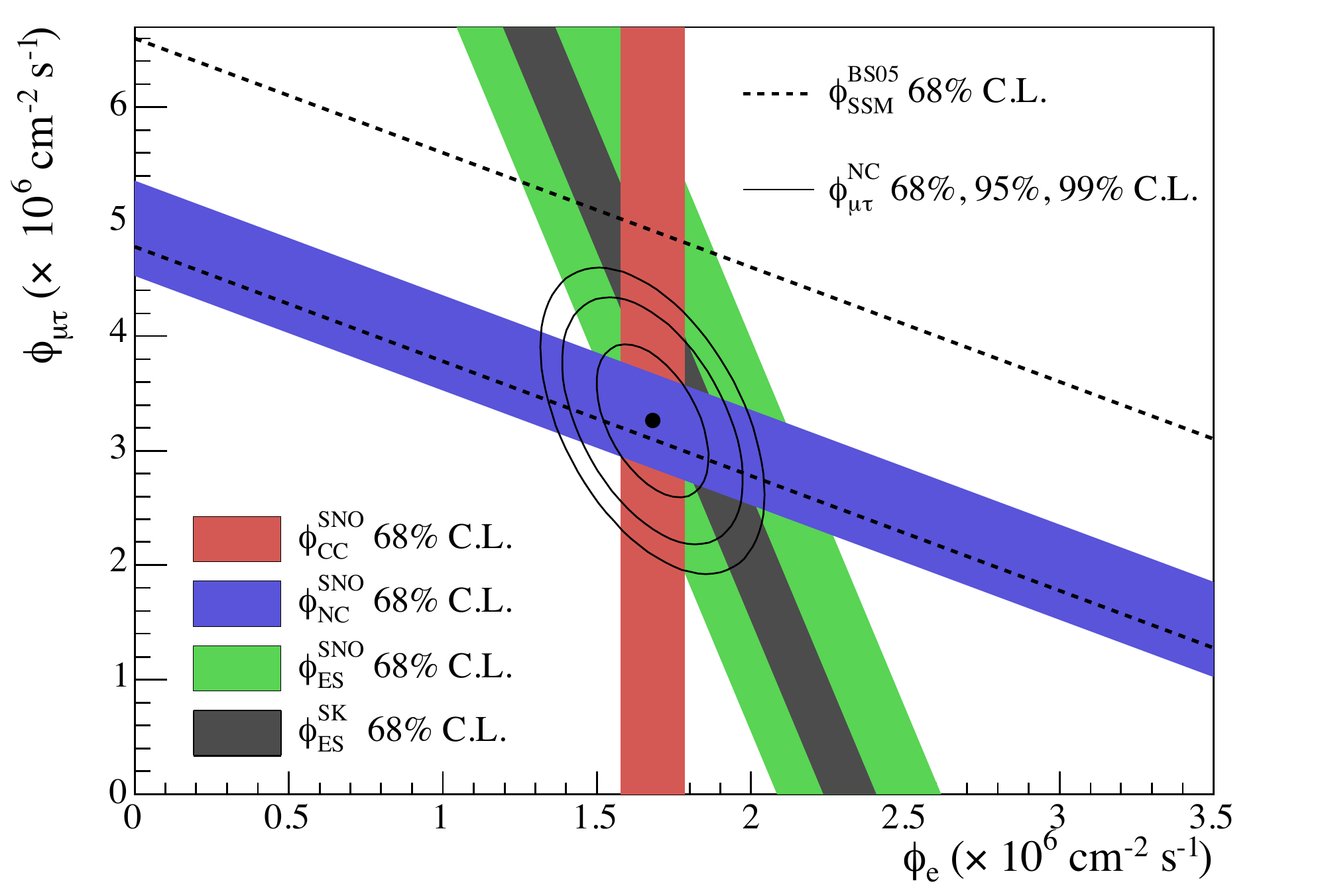}
\caption{The SNO CC, ES, and NC and Super-Kamiokande ES constraints on the $\nu_e$ and 
$\nu_\mathrm{\mu \tau}$ 
components of the $^8$B flux are shown.  The SSM total flux is also shown.  All results are all
consistent with a terrestrial flux of solar neutrinos that is about one-third $\nu_e$s and
two-thirds $\nu_\mathrm{\mu \tau}$ \cite{Aharmim2012}. Reproduced with permission from
\cite{SNOsalt}.}
\label{fig:SNO}
\end{center}
\end{figure}

Results from the SNO detector were definitive in proving new particle physics was responsible for the solar
neutrino deficit Ray Davis first discovered.  Like Super-Kamiokande, SNO was able to detect elastic scattering (ES)
events
\begin{equation}
\label{sno1}
\nu_x + e^- \rightarrow \nu_x + e^-
\end{equation}
which, due to the charged-current contribution to $\nu_e$ scattering, has a relative sensitivity to
$\nu_e$s and heavy-flavor neutrinos of $\sim$ 7:1.  In addition, 
SNO could detect $\nu_e$s through the charged
current (CC) reaction 
\begin{equation}
\label{sno2}
\nu_e + d \rightarrow p + p + e^-,
\end{equation}
by observing the Cherenkov light from the recoiling electron.  Unlike the ES case, in the CC
reaction the $e^-$  carries away most of the energy release, so that the electron recoil spectrum becomes an
important test of the neutrino spectrum. SNO's third reaction was the neutral current (NC) breakup reaction
\begin{equation}
\label{sno3}
 \nu_x (\overline{\nu}_x)+ d \rightarrow \nu_x (\overline{\nu}_x)+
p + n.  
\end{equation}
This reaction is independent of flavor and thus sensitive to the total flux of neutrinos: by detecting the neutron,
one can determine a rate that corresponds to an integral over the neutrino spectrum above the 2.22 MeV
deuterium breakup threshold. To exploit this reaction, great care had to be taken to reduce background sources
of neutrons from natural radioactivity and from cosmic ray muons.  SNO's great depth, two kilometers 
below the surface, was helpful in addressing the latter.  The SNO experiment was carried out in three phases
in which distinct methods were used in the NC channel, neutron capture on deuterium,
on Cl introduced as salt in the heavy water, and in an array of $^3$He proportional counters 
constructed in the central detector.  The SNO results for the CC, ES, and NC channels,
the Super-Kamiokande ES results, and the predictions of the SSM circa 2005 are shown
in Fig. \ref{fig:SNO}.  The data are consistent,
showing that approximately two-thirds of the solar neutrinos arriving at earth are heavy-flavor neutrinos \cite{Ahmad:2001an,
Aharmim2012}.  The
total flux is consistent with the SSM, given assigned uncertainties.

Table \ref{tab:2} summarizes the properties of the solar neutrino experiments carried out to date.

\begin{table}
\begin{center}
\begin{minipage}[t]{16.5 cm}
\caption{Solar neutrino experiments.}
\label{tab:2}
\end{minipage}
\begin{tabular}{|c|c|c|c|c|c|}
\hline {\em Experiment} & {\em Type} & {\em Method} & {\em Dates} & {\em Threshold(MeV)}  & {\em Main Fluxes} \\
\hline  \hline
Homestake Cl   & radiochem & CC & 1968-2002  & 0.811 & $^8$B, $^7$Be \\
\hline Kamiokande  & active & ES & 1987-1995 & 7.0-9.0 &  $^8$B \\ 
\hline SAGE   & radiochem & CC  & 1990-present &  0.233   & pp, $^7$Be \\
\hline GALLEX/GNO   & radiochem &  CC &1991-2003 &  0.233  & pp, $^7$Be \\
\hline Super-Kamiokande  & active & ES & 1996-present &  4.0-7.0   & $^8$B \\ 
\hline SNO & active & ES/CC/NC  & 1999-2006 & 4.0-7.25; 2.22(NC)  & $^8$B \\
\hline Borexino & active & ES  & 2008-present & $\sim$ 0.8   &$^7$Be, pep, $^8$B \\  \hline
\end{tabular}
\end{center}
\end{table}

The solar neutrino demonstration of neutrino oscillations was confirmed in the KamLAND reactor neutrino experiment \cite{Abe:2008aa}. The KamLAND experiment used neutrinos from 55 Japanese nuclear reactors with a flux-weighted average baseline of 180 km, a distance fortuitously appropriate for observing oscillations governed by the solar  $\delta m_{21}^2$.  The KamLAND results provided a precise determination of $\delta m_{12}^2
\sim (7.41^{+0.21}_{-0.19}) \times 10^{-5}$ eV$^2$, eliminating
an ambiguity that existed in analyses that included only solar neutrino data: both the
LMA (large mixing angle) solution corresponding to $\delta m_{12} \sim 8 \times 10^{-5}$ eV$^2$ and
the LOW solution, $\delta m_{12}^2 \sim 10^{-7}$ eV$^2$, were allowed at 67\% c.l., prior to
KamLAND measurements. KamLAND results eliminated the LOW solution.

Neutrino propagation in the Sun is influenced by matter, which alters the relationship between the local mass 
eigenstates, effectively rotating the local mixing angle toward $\pi/2$ with increasing density.
The electron neutrino survival probability at Earth, averaged over the solar core production region (which removes interference terms between the local heavy and light mass eigenstates), is



\begin{equation}
\label{day}
P_{2\times 2}^D( \nu_e \rightarrow \nu_e) = \frac{1}{2} +\frac{1}{2} \cos 2 \theta_{12} \langle \cos 2 \theta_i 
\rangle_{source}  \left( 1 - 2 P_{hop} \right) .
\end{equation}
Here $\langle \cos{2 \theta_i} \rangle_{source}$ is the local mixing angle at the point of neutrino production,
suitably averaged over the neutrino-producing region of the solar core; the superscript $D$ indicates 
that the neutrino has been observed during the day, so that
matter effects associated with passage through the earth play no role; and the subscript $2 \times 2$ 
indicates that subdominant oscillations involving the third flavor have been neglected.
$P_{hop}$ is the probability that the neutrino hops from the heavy-mass trajectory to the light-mass
trajectory on traversing the level-crossing point \cite{Haxton:1986dm,Parke:1986jy}.  This accounts for possible 
non-adiabatic behavior: if the density in this region is treated linearly, one finds
\begin{equation} P_{hop} = e^{ - \pi \gamma_c/2} 
\end{equation}
where $\gamma_c$ is proportional to the ratio of the Sun's density scale
height to the local oscillation length at the crossing point.  (Thus large $\gamma_c$ corresponds to small
changes in the solar density over an oscillation length, the adiabatic limit where $P_{hop} \rightarrow 0$.)

Solar neutrinos detected at night go through the Earth, hence they would feel additional matter effects \cite{Bahcall:1997jc}. 
The night-time solar electron neutrino survival probability, averaged over the Earth-Sun distance, is given by
\cite{GonzalezGarcia:2000ve,Minakata:2010be} 
\begin{equation}
\label{night}
P_{2\times2}^N( \nu_e \rightarrow \nu_e) = \frac{1}{2} +\frac{1}{2} \langle \cos 2 \theta_i \rangle_{source}  \left( 1 - 2 P_{hop} \right) \cos 2 \theta_e \cos 2(\theta_e - \theta_{12}) 
\end{equation}
where $\theta_e$ is the matter mixing angle inside the Earth. In writing Eq. (\ref{night}), matter density in the mantle of the Earth is assumed to be constant. The day-night asymmetry $A$ is defined as
\begin{equation}
\label{14}
\frac{A}{2} = \frac{P^N-P^D}{P^N + P^D}. 
\end{equation}
The terrestrial matter effects on solar neutrinos that can generate day/night differences
are quite small: based on current neutrino parameters obtained from
global analyses, a $\sim$ 3\% day-night effect is expected in ES measurements of the $^8$B spectrum.  
The latest available Super-Kamiokande results 
remain consistent with no effect at 2.6 $\sigma$ \cite{Smy2012}, though it is quite possible that with the data now being
accumulated in Super-Kamiokande IV low-threshold running, this conclusion may change.
The corresponding SNO result  is less restrictive, consistent with the null hypothesis of no day/night effects
at 61\% c.l. \cite{Aharmim2012}.   The predicted day/night effect in Borexino is tiny, 
consistent with the collaboration's determination A = 0.001 $\pm$ 0.012 (stat) $\pm$ 0.007 (syst) \cite{Bellini2012b}.
(The absence of day/night effects in Borexino's $^7$Be flux measurement is an alternative means of
eliminating the LOW solution.)

\begin{figure}
\begin{center}
\includegraphics[width=14cm]{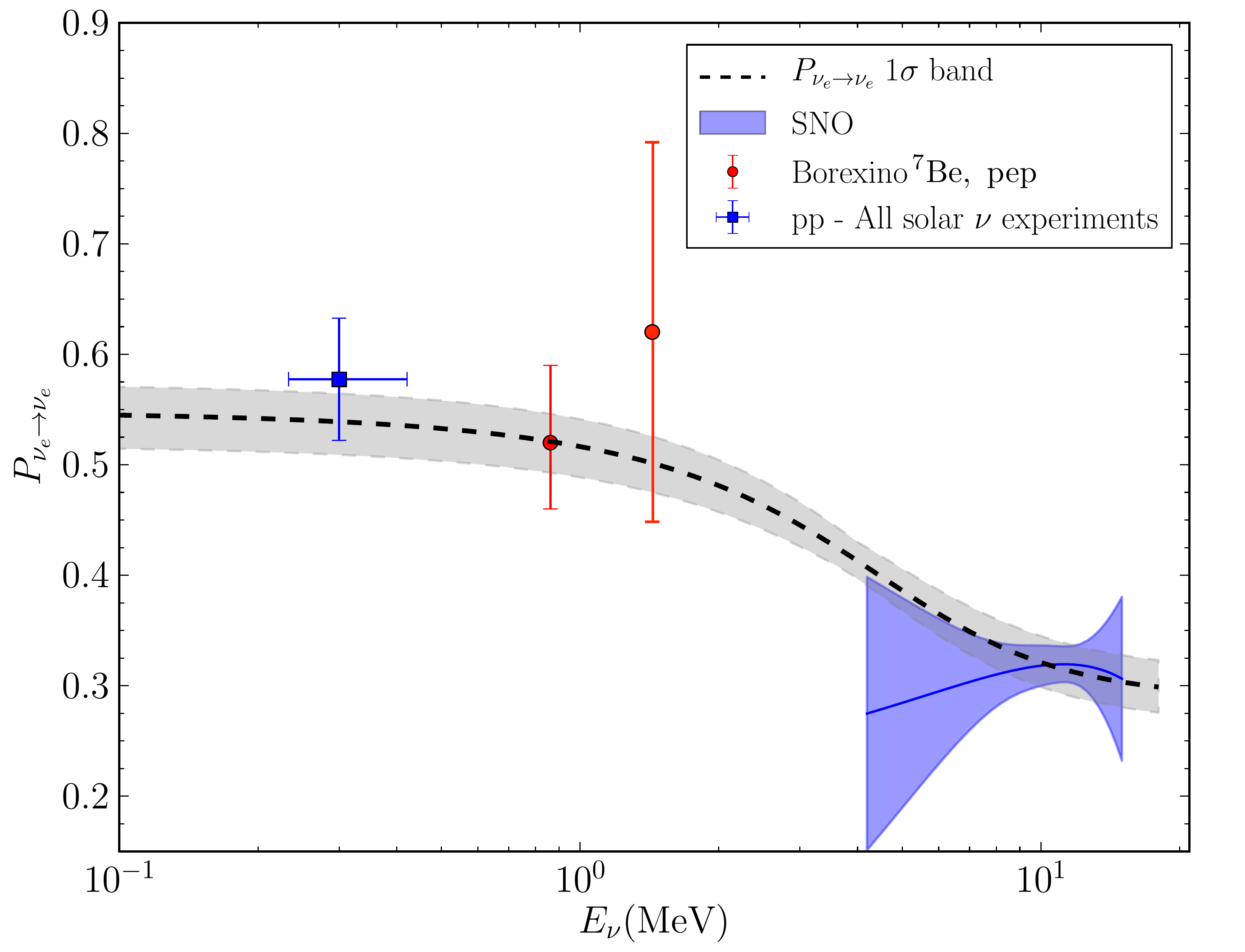}
\caption{Global analysis pp, Borexino $^7$Be and pep, and SNO LETA $^8$B deduced survival probabilities
$P_{\nu_e}$.  Figure from \cite{Robertson:2012ib}, which was adapted from \cite{Aharmim2012,Bellini2012}.}
\label{fig:distort}
\end{center}
\end{figure}

In contrast, matter effects associated with passage
through the Sun are large, with the additional suppression being substantial for the higher
energy $^8$B neutrinos.  Because matter effects either increase of decrease the local splitting between the
mass eigenstates, depending on the hierarchy, solar matter effects allow one to 
establish that $\delta m_{21}^2$ is positive ($m_2 > m_1$).
Consequently $B$ of Eq. (\ref{4a}) is positive.   When one uses neutrino oscillation
parameters taken from current global analyses, one finds that low-energy solar neutrino fluxes
(pp, pep, $^7$Be) reside in a region where vacuum effects are dominant,
\begin{equation}
 P_{\nu_e}^\mathrm{vaccum} \sim 1 -{1 \over 2} \sin^2{2 \theta_{12}} \sim 0.57
 \label{eq:vac}
 \end{equation}
 while oscillation probabilities for the high-energy portion of the $^8$B neutrino spectrum measured by SNO and Super-Kamiokande 
 are substantially affected by matter
 \begin{equation}
 P_{\nu_e}^\mathrm{high~density} \rightarrow \sin^2{\theta_{12}} \sim 0.31.
 \label{eq:mat}
 \end{equation}
 These expectations are clearly apparent in the data:  the pp flux determined in luminosity-constrained
 global analyses and the $^7$Be and pep fluxes determined by Borexino \cite{Bellini2011,Bellini2012}
 all show the expected turn-up in $P_{\nu_e}$ at low $E_\nu$. (See Fig. \ref{fig:distort}.)
 
 One of the goals of both SNO (CC/ES) and Super-Kamiokande (ES) has been to see the matter effects imprinted on the
 scattered electrons.  SNO's low-energy-threshhold  (LET) analysis \cite{Aharmim2012} and
 Super-Kamiokande IV running both employ total-energy thresholds of $\sim$ 4.0 MeV.  This provides
 sensitivity to a region of the spectrum where the matter-to-vacuum transition might become apparent.
 The $P_{\nu_e}$s deduced from the SNO LET
 analysis are included in Fig. \ref{fig:distort}.  While remaining consistent with global
 analysis predictions, the best fit is trending away from expectations.  
 
The three-flavor analysis
based on all solar data supplemented by KamLAND  yields \cite{Aharmim2012}
\begin{equation}
\sin^2{\theta_{12}} = 0.308 \pm 0.014~~~\delta m_{21}^2 \sim (7.41^{+0.21}_{-0.19}) \times 10^{-5} \mathrm{eV}^2~~~\sin^2{\theta_{13}} = 0.025^{+0.018}_{-0.015} 
\end{equation}
These results are in excellent agreement with the global analysis that takes into account all neutrino data,
including beam experiments and new reactor experiments.  The Bari \cite{Bari} and Valencia \cite{Valencia} analyses yield, respectively,
\begin{equation}
\sin^2{\theta_{12}} = \Bigg\{\begin{array}{c} 0.306^{+ 0.018}_{-0.016} \\ \\0.319^{+ 0.016}_{-0.018} \end{array}
~~~~~~\delta m_{21}^2 \sim \Bigg\{  \begin{array}{c}  (7.54^{+0.26}_{-0.22}) \\  \\(7.62^{+0.19}_{-0.19}) \end{array}  \times 10^{-5} \mathrm{eV}^2~~~~~~\sin^2{\theta_{13}} =\Bigg\{ \begin{array}{c}  0.0243^{+0.0025}_{-0.0027} \\ \\  0.0248^{+0.0028}_{-0.0030} 
 \end{array}
\end{equation}
The solar neutrino analysis supplemented by KamLAND produces  a best-value estimate of $\theta_{13}$
that is in very good agreement with the Bari and Valencia results.   However, the global analyses provide a more certain
determination of $\theta_{13}$: this reflects the impact of the recent 
Daya Bay, RENO, and Double Chooz reactor experiments, which tightened the error bars
on $\theta_{13}$ by almost an order of magnitude.  We 
discuss these experiments in Sec. \ref{sec:theta13}.

The original goal of solar neutrino spectroscopy was to use neutrinos to test the hypothesis that the Sun
is powered by the pp chain and to measure the solar core temperature to high accuracy.  This program was
delayed by the long effort to identify the origin of the solar neutrino problem and then to accurately determine the parameters
governing neutrino oscillations.  Consequently today, given the few-percent determinations made of the relevant
mixing angles and $\delta m_{12}^2$, it has finally become possible
 to use neutrinos as quantitative astrophysical probes of the Sun and the SSM.  In addition to weak
 interaction uncertainties, the SSM depends on the
 accuracy with which an additional $\sim$ 20 input parameters are known.  These parameters are associated with 
 SSM's nuclear and atomic
physics  and with properties such as the Sun's  initial metallicity, mass, age, luminosity, and the diffusion coefficient.
Decades of laboratory and theoretical work on the nuclear and atomic physics input has helped
make the SSM a predictive model, where output uncertainties are known.  The laboratory astrophysics of the SSM
 was advanced by community efforts like that leading to the analysis of cross sections and S-factors
published in 1998 \cite{Adelberger:1998qm} (Solar Fusion I).  The key uncertainties identified in that analysis led 
to a decade of
further work on critical cross sections, such as that for $^7$Be(p,$\gamma)$, and to the recent re-evaluation
of cross sections summarized in \cite{Adelberger:2010qa} (Solar Fusion II).  Current SSM predictions of
neutrino and helioseismic observables have uncertainties based on evaluated input
laboratory data.  Discrepancies in SSM predictions that significantly exceed model uncertainties
could indicate problems with the underlying astrophysics assumptions of the SSM.

The key discrepancy in current SSM predictions is associated with its initial metallicity (abundances of 
elements heavier than hydrogen and helium).   The overall solar metallicity scale is usually determined from
analyses of solar photo-absorption lines.   Recently more realistic models of the photosphere were 
constructed to account for its dimensionality and granularity, effects ignored in traditional 1D slab models.
When applied to the photospheric data, these new models yield an abundance ratio of metals to hydrogen $(Z/X)_{\odot} = 0.0178$   \cite{Asplund:2009fu}, compared to the previously accepted value of $(Z/X)_{\odot} = 0.0229$ \cite{Grevesse:1998bj}. Recent SSM calculations have been carried out for both metallicities \cite{serenelli}, while
also incorporating the new results from Solar Fusion II.  The helioseismic predictions of the low-metallicity
SSM are in significant disagreement with observation.  The neutrino flux predictions of both models
are in reasonable agreement with the data, though it is apparent that a solar core metallicity set to the average of
the low- and high-metallicity models would improve the agreement.

This has opened up two directions of study.  First, one of the central assumptions of the SSM is that the
Sun was homogeneous at the start of the main sequence.  There is little observational support for this
assumption, given that it is known, late in the evolution of the solar disk, that the process of planetary
formation scoured 50-90 $M_\oplus$ of excess metal from the disk.  New solar models have
been built to take account of the possible accretion of metal-depleted gas onto the proto-Sun, late
in its evolution when such accretion would alter convective-zone abundances but not affect the core.
It has been found that both solar neutrino and helioseismic data severely constrain such accreting models \cite{serenelli}.
Second, the volatile elements CNO play a key role in the solar abundance problem.  The SSM produces about
1\% of its energy through the CN cycle, yielding significant fluxes of $^{13}$N and $^{15}$O solar
neutrinos.  It has been shown, because of the CN cycle's additional linear dependence on the abundance
of C+N (these metals catalyze the proton burning), that the
C+N abundance of the solar core could be determined to an accuracy of about 10\% from the CN neutrino fluxes \cite{aswch}.  
This would be very significant given the current 30+\% discrepancy between the low- and high-metallicity
SSMs.  The CN neutrinos can be measured
in very deep scintillation detectors such as SNO+, and possibly also in shallower detectors such as
Borexino, if strategies for suppressing cosmic-ray muon backgrounds can be further developed.

There are other gaps in our program of solar neutrino spectroscopy.  While the pp flux is tightly constrained
by the solar luminosity constraint, given existing $^7$Be and $^8$B neutrino flux measurements,
this constraint itself should be checked:  are the Sun's neutrino and photon lumninosities consistent,
as predicted in steady-state solar models?
Recently the Borexino collaboration succeeded in detecting the pep neutrinos \cite{Bellini2012},
though the accuracy of the measurement is so far limited.  This flux should track that of the low-energy pp neutrinos,
the dominant solar neutrino flux arising directly from p+p fusion.  While the pp flux is constrained by the
results from the Ga experiments, no real-time detection has been made. The relation between the photon and
neutrino luminosities in steady-state models is \cite{Spiro:1990vi}
\begin{equation}
\label{lum}
\frac{L_{\odot}}{4\pi d^2} =\sum_i \left( Q - \langle E \rangle
\right)_i \phi_i, 
\end{equation} 
where $d$ is the average Sun-Earth distance (1 a.u.), $Q$ is the energy released in the nuclear reactions, $\langle E
\rangle$ is the average energy loss by neutrinos, and $\phi_i$ is the
neutrino flux at earth, coming from the reaction of type $i$. (There are corrections to this equation due to the necessary 
inequalities between different fluxes to take into account the order in which nuclear reactions take place \cite{Bahcall:1995rs,HR1996}). 
A test of this relation would require a precise measurement of both the pp- and CNO-chain neutrino fluxes. Such a test
would place stringent limits on new neutrino species, while also checking the assumption
of steady-state solar burning.

\section{Measuring $\theta_{23}$: Atmospheric and Accelerator Neutrinos}

The hadronic cosmic ray flux incident on Earth is dominated by energetic protons with fewer numbers of heavy nuclei. 
When the primary cosmic rays collide with the oxygen and nitrogen in Earth's atmosphere they produce secondary pions, kaons,
and muons. Atmospheric neutrinos come from the decay of these secondaries.  For energies less than 1
GeV all these secondaries decay in the atmosphere
\begin{eqnarray}
\pi^{\pm} (K^{\pm}) &\rightarrow &\mu^{\pm} + \nu_{\mu}
(\overline{\nu}_{\mu}), \nonumber\\ \mu^{\pm} &\rightarrow & e^{\pm} +
\nu_e (\overline{\nu}_e) +  \overline{\nu}_{\mu} (\nu_{\mu}).
\end{eqnarray}
Consequently one expects the ratio

\begin{equation}
r = (\nu_e + \overline{\nu}_e) / (\nu_{\mu} + \overline{\nu}_{\mu})
\end{equation}

\noindent
to be approximately 0.5 in this energy range.  This relationship can be checked as a function of distance up
to $\sim$ 13000 km, by observing atmospheric neutrinos in a deep detector as a function of zenith angle.  
It has been customary to use a ratio-of-ratios
\begin{equation}
R = { \left[ (\nu_e + \overline{\nu}_e) / (\nu_{\mu} + \overline{\nu}_{\mu}) \right]_\mathrm{data} \over  \left[ (\nu_e + \overline{\nu}_e) / (\nu_{\mu} + \overline{\nu}_{\mu}) \right]_\mathrm{Monte~Carlo}  }
\end{equation}
in this check, normalization the measured $r$ to a Monte Carlo calculation of the ratio assuming 
no oscillations but including corrections such as geomagnetic effects on the cosmic ray flux. 

Early atmospheric neutrino experiments were inconsistent with one another, with some finding
$R \sim 1$  while others finding diminished values. 
Significant indications of an anomaly came from the IMB \cite{IMB} and Kamiokande \cite{KamAtmos}
detectors.  IMB first noticed a possible deficit of neutrino-induced muon events in 1986, while
Kamiokande established a deficit in excess of 4$\sigma$ by 1988.  By 1998 the anomaly was also
apparent in data taken from the Soudan detector and from Super-Kamiokande \cite{Fukuda:1998mi}. 

In early experiments it was difficult to adequately explore $R$ because the lack of statistics prevented a detailed
analysis as a function of zenith angle.  This changed with the construction of Super-Kamiokande, which
provided a fiducial volume of $\sim$ 20 ktons.  In an initial analysis the Super-Kamiokande Collaboration  found
\begin{equation}
R = \Bigg\{ \begin{array}{ll} 0.61 \pm 0.03 (\mathrm{stat}) \pm 0.05 (\mathrm{sys}) &~~~ \mathrm{sub-GeV~events,~fully~contained} \\
0.66 \pm 0.05 (\mathrm{stat}) \pm 0.08 (\mathrm{sys}) &~~~ \mathrm{multi-GeV~events,~fully~or~partially~contained} \end{array}
\end{equation}
The 1998 Super-Kamiokande analysis, based on 33 kton-years of data, showed a zenith-angle dependence
for $R$ inconsistent with theoretical calculations of the atmospheric flux, in the absence of oscillations. The
distance-dependent muon deficit was consistent with oscillations governed by 
neutrino mass differences within the range $5 \times 10^{-4} \mathrm{eV}^2 < |\delta m_{32}^2 | <
6 \times 10^{-3} \mathrm{eV}^2$ \cite{Fukuda:1998mi}.  This differed from that later determined for solar neutrinos.  The collaboration
concluded that the measurements were consistent with the two-flavor oscillation $\nu_\mu \rightarrow \nu_\tau$, 
providing evidence for massive neutrinos \cite{Fukuda:1998mi}.

Super-Kamiokande I collected $\sim$ 15,000 atmospheric neutrino events over approximately five years
of running.  The collaboration's zenith-angle analysis of the data placed the first oscillation minimum at
$L/E \sim$ 500 km/GeV, so that $L_0 \sim$ 1000 km for a 1 GeV muon neutrino.  Results based on 3903 days
of data from Super-Kamiokande I-IV, including 1097 days from the current phase IV, were incorporated in a
three-flavor analysis in which $\sin^2{\theta_{13}}$ was set to 0.025, based on Daya Bay, RENO, and Double
Chooz results.  The analysis yielded \cite{Itow, Ashie:2005ik}
\begin{eqnarray} 
\mathrm{Normal~hierarchy:}&& \begin{array}{lrr} \delta m_{32}^2 = & (2.66^{+0.15}_{-0.40} ) \times 10^{-3} \mathrm{eV}^2 & (1 \sigma) \\ \sin^2{\theta_{23}} = & 0.425^{+0.194}_{-0.034} & (90\% \mathrm{c.l.}) \end{array} \nonumber \\
\mathrm{Inverted~hierarchy:}&& \begin{array}{lrr} \delta m_{32}^2 = & (2.66^{+0.17}_{-0.23} ) \times 10^{-3} \mathrm{eV}^2 & (1 \sigma) \\ \sin^2{\theta_{23}} = & 0.575^{+0.055}_{-0.182} & (90\% \mathrm{c.l.}) \end{array} \nonumber \
\end{eqnarray}
The significant difference between the normal- and inverted-hierarchy best values for $\sin^2{\theta_{23}}$ reflects the
flatness of the $\chi^2$ fit around $\sin^2{\theta_{23}} \sim 0.5$, with shallow local minima appearing above and below 
this value.

%
%
Accelerator experiments were conducted by the K2K \cite{Ahn:2006zza} and MINOS \cite{Adamson:2012rm,Adamson:2012gt}
collaborations, with baselines designed to be sensitive to the $\delta m_{32}^2$ scale probed with atmospheric neutrinos.
The MINOS beam-neutrino analysis, which includes atmospheric neutrino data from that detector,
determined $|\delta m_{32}^2| = 2.39^{+0.09}_{-0.10}$ eV$^2$ \cite{Nichol}, a value consistent with the somewhat
broader range allowed by the atmospheric results.  
The sign of $\delta m_{32}^2$ has not been determined, though one can see from the results above that there
is some sensitivity to the hierarchy in atmospheric neutrino fits.  There are several pending proposals to determine the hierarchy:  new, long-baseline neutrino beam experiments (some of these depend on joint 
analyses of beam and atmospheric neutrino data), high-density arrays to enhance the sensitivity of large
detectors like IceCube to atmospheric neutrinos, and new reactor-neutrino experiments with
very large detectors to increase sensitivity to subdominant oscillations.

\section{Measuring $\theta_{13}$: Reactor and Accelerator Neutrinos}
\label{sec:theta13}


Initial comparisons of the solar neutrino  and KamLAND reactor neutrino experiments revealed hints of a non-zero
value of $\theta_{13}$ \cite{Balantekin:2008zm,Fogli:2008jx}.   The solar neutrino+KamLAND analysis performed by
the KamLAND Collaboration
in 2010 yielded $\sin^2{\theta_{13}} = 0.020^{+0.16}_{-0.16}$ \cite{Gando}, with a nonzero value suggested
at 79\% c.l.  Indications of electron neutrino appearances from accelerator-produced off-axis muon neutrinos in the T2K \cite{Abe:2011sj} and MINOS \cite{Adamson:2009yc} experiments also suggested relatively large values of $\theta_{13}$. 
In 2012 the Double Chooz reactor collaboration announced data that suggested a nonzero value for $\sin^2{2 \theta_{13}}$
at 94.6\% c.l. \cite{Abe:2012tg}, followed in quick succession by definitive results from the Daya Bay \cite{An:2012eh}
and RENO collaborations \cite{Ahn:2012nd} . The various
determinations of $\sin^2 2 \theta_{13}$ are summarized in Table \ref{tab:3}.

\begin{table}
\begin{center}
\begin{minipage}[t]{16.5 cm}
\caption{Values of $\sin^2 2 \theta_{13}$ measured at various experiments.  The value quoted for the MINOS experiment corresponds to $2 \sin^2\theta_{23} \sin^2 2 \theta_{13}$. }
\label{tab:3}
\end{minipage}
\begin{tabular}{lr}
\hline
{\bf Experiment}
  & {\bf Measured ${\mathbf \sin^2 2 \theta_{13}}$} \\
\hline
Daya Bay \cite{An:2012bu}  & $0.089\pm 0.010({\rm stat})\pm0.005({\rm syst})$\\
RENO \cite{Ahn:2012nd}  & $0.113 \pm 0.013 ({\rm stat}) \pm 0.019 ({\rm syst})$\\
Double Chooz \cite{Abe:2012tg} &    $0.109 \pm 0.030 ({\rm stat}) \pm 0.025 ({\rm syst})$       \\
T2K (assuming $\theta_{23} = \pi/4$) \cite{t2k} & $0.104^{+0.060}_{-0.045} $\\
MINOS \cite{Adamson:2011qu} (see note)  & $+0.041^{-0.047}_{+0.031}$ (normal) $0.079^{+0.071}_{-0.053}$ (inverted)\\
\hline
\end{tabular}
\end{center}
\end{table}

Reactor experiments are disappearance experiments, with $\theta_{13}$ determined from the magnitude
of the electron antineutrino flux loss with distance.  The disappearance probability, taken from Eq. (\ref{ee}), is 
\begin{equation}
\label{ex}
1- P (\nu_e \rightarrow \nu_e) = \sin^2 2 \theta_{13} \left[ \cos^2 \theta_{12} \sin^2( \Delta_{31} L) + \sin^2 \theta_{12} \sin^2 (\Delta_{32} L) 
\right] + \cos^4 \theta_{13} \sin^2 2 \theta_{12} \sin^2 (\Delta_{21} L) .
\end{equation}
If the experiment is done close to the reactor, $\Delta_{21} L \sim 0$ so that the ``solar" oscillation represented by the second term can be ignored. Using the fact $\Delta_{32}  \sim \Delta_{31} $, suggested by the solar neutrino experiments, one obtains the disappearance probability
\begin{equation}
\label{11}
1- P (\nu_e \rightarrow \nu_e) = \sin^2 2 \theta_{13} \sin^2( \Delta_{31} L) . 
\end{equation}
Consequently the very short-baseline reactor neutrino experiments, such as Daya Bay, Double Chooz, and RENO, unambiguously measure $\theta_{13}$, requiring no knowledge of other neutrino mixing angles or the mass hierarchy. Furthermore, the multiple detector configurations currently employed in the Daya Bay and RENO experiments also 
minimize the effects of reactor neutrino spectrum uncertainties.

The situation is rather different for longer-baseline experiments searching for the appearance of electron neutrinos in a flux of muon neutrinos. 
For longer baselines neutrinos travel trough the Earth, hence matter effects need to be included. The resulting rather complicated expressions, which depend on many of the neutrino parameters, can be calculated in a series expansion \cite{Akhmedov:2004ny}. 
The appropriate appearance probability in the lowest order is given by 
\begin{equation}
\label{xe}
P (\nu_{\mu} \rightarrow \nu_e) \sim \frac{ \sin^2 2 \theta_{13} \sin^2 \theta_{23}}{(1- G_F N_e /\sqrt{2} \Delta_{31})^2}  
\sin^2 \left[ \left ( \Delta_{31} - \frac{G_F N_e}{\sqrt{2}} \right) L \right] + {\cal O} (g) ,
\end{equation}
where 
\begin{equation}
\label{gdef}
g = \frac{\delta m_{21}^2}{\delta m_{31}^2} \sim 0.03 .
\end{equation}
The next order correction in $g$ to the electron neutrino appearance probability 
brings in an additional dependence on the CP-violating phase in the neutrino mixing matrix. The denominator of the term multiplying the oscillating term in Eq. (\ref{xe}) depends on the sign of $\delta m^2$, i.e. the mass hierarchy of the neutrinos. 
Consequently, appearance experiments such as T2K \cite{t2k} and MINOS \cite{Adamson:2011qu} cannot disentangle $\theta_{13}$ from other observables such as the mass hierarchy, $\theta_{23}$, and the CP-violating phase. 
Note that an intermediate-baseline ($L \sim 60$ km) reactor antineutrino experiment will also have some sensitivity to the neutrino mass hierarchy 
\cite{Batygov:2008ku}.

\section{Final Remarks}
 
The puzzle that arose from early solar neutrino measurements had an interesting resolution, still incomplete.
 We have reviewed what has been learned in the past 15 years about the properties of neutrinos from 
oscillations in vacuum and in matter, using astrophysical neutrino sources and terrestrial beams.
A great deal of progress has been made, with three mixing angles and two mass differences now accurately
determined, and with the sign of $\delta m_{12}^2$ deduced the effects of solar matter on solar neutrino oscillations.
Yet there are several important properties of the light neutrinos that we have not yet been able
to determine: the overall scale of the neutrino masses, the mass hierarchy, the Dirac CP-violating phase in the mixing matrix, and neutrino charge-conjugation properties.  We have the tools to make further progress.  Long-baseline 
neutrino beam oscillation experiments could distinguish between the normal and inverted mass patterns and determine the
CP phase.  As we have noted, there may be alternative strategies in which
reactor and atmospheric neutrino sources are exploited to probe the neutrino mass
hierarchy.  A variety of cosmological tests are sensitive to the absolute scale of neutrino mass:
the distinctive imprint of neutrino mass on the growth of large-scale structure, affecting both the scale and red-shift
dependence of that growth, produces changes in the power spectrum that are roughly an order of magnitude
larger than one might naively guess, based on the contribution of neutrinos to the mass density.  There are a
variety of ongoing projects to improve laboratory sensitivity to neutrino mass, such as the 
high-sensitivity spectrometer KATRIN for measuring tritium beta decay
and the MARE micro-calorimeters for measuring $^{187}$Re beta decay, as well as new technologies
for overcoming resolution challenges, such as the Project 8 extension to KATRIN.  Finally, the exciting question
of whether Nature exploits the neutrino's unique Majorana mass mechanism is stimulating next-generation
efforts in double beta decay.  As progress is made on all of these fronts, neutrino oscillations will remain
one of our most important and most versatile tools from probing the new properties of neutrinos, and thus for seeking
new physics beyond the SM.
 
This work of ABB was supported in part 
by the U.S. National Science Foundation Grant No.  PHY-1205024 (Wisconsin) and the 
University of Wisconsin Research Committee through funds
granted by the Wisconsin Alumni Research Foundation.  The work of WCH was supported in part by
the US DOE under DE-SC00046548 (UC Berkeley) and DE-AC02-98CH10886 (LBL),
and by the UC Office of the President.

\end{document}